\documentclass{elsart}

\usepackage{amsmath,amssymb,graphicx}

\def\un{{\underline 1}}

\def\kvec{\mbox{${\rm\bf k}$}}
\def\kzt{\mbox{${\tilde k}_z$}}

\def\kvece{\mbox{$\rm\bf k_e$}}
\def\betat{\mbox{${\tilde\beta}$}}
\def\kveco{\mbox{$\rm\bf k_o$}}
\def\tbyt{\mbox{$2\times 2$ }}
\def\fbyf{\mbox{$4\times 4$ }}

\def\Ev{\mbox{${\rm\bf E}$}}
\def\Evo{\mbox{${\rm\bf E_{out}}$}}
\def\Evoo{\mbox{${\rm\bf E_{out}^{(0)}}$}}
\def\Evi{\mbox{${\rm\bf E_{in}}$}}
\def\Hv{\mbox{${\rm\bf H}$}}
\def\evec{\mbox{${\bf e}$}}
\def\ovec{\mbox{${\bf o}$}}
\def\khat{\mbox{${\rm\bf\hat k}$}}
\def\phat{\mbox{${\rm\bf\hat p}$}}
\def\shat{\mbox{${\rm\bf\hat s}$}}
\def\vhat{\mbox{${\rm\bf\hat v}$}}

\def\er{\mbox{ ${\rm\bf r}$}}

\def\xhat{\mbox{${\rm\bf\hat x}$}}
\def\yhat{\mbox{${\rm\bf\hat y}$}}
\def\zhat{\mbox{${\rm\bf\hat z}$}}

\begin{document}

\begin{flushright}
{\bf LAL 05-21}\\
\vspace*{0.1cm}
{ May 2005}\\
\end{flushright}

\begin{frontmatter}

\title{Transmission matrix of a uniaxial optically active crystal platelet }

\author{F. Zomer}
 \ead{zomer@lal.in2p3.fr}

\address{ Laboratoire de l'Acc\'el\'erateur Lin\'eaire, Institut National
 de Physique Nucl\'eaire et de
 Physique des Particules - Centre National de la Recherche Scientifique
 et Universit\'e de Paris-Sud, B.P. 34 - 91898 Orsay cedex, France.} 

\begin{abstract}
Expressions corresponding to the transmission of
 a uniaxial optically active crystal platelet are provided for an optical axis
 parallel and perpendicular to the plane of interface. The 
 optical activity is taken into account by a consistent multipolar expansion of 
the
 crystal medium response due to the path of an electromagnetic wave. Numerical
 examples of the effect of the optical activity are given
 for quartz platelets of chosen thicknesses. The optical activity's effects
 on the variations of the transmission 
 of quartz platelets as a function of the angle of incidence is also 
investigated.
\end{abstract}


\end{frontmatter}

\section{Introduction}

Optical activity in crystals is a phenomenon which has been known for a long 
time 
(see Ref. \cite{condon} for an historical review).
 Two methods are currently used to describe the effect of the crystal optical 
activity
 on plane wave propagation. The first one 
 is phenomenological and is valid under normal incidence
 \cite{ny,scierski,kazi,kush,chou,hernandez}. The second one 
 is a first principle method and consists in solving the Maxwell equations
 with modified constitutive equations \cite{condon,yariv,silverman,miteva}
 ({\it i.e.} the relationship between the electric
 and magnetic fields vectors and the electric displacement and magnetic 
induction field vectors).
 The choice of the constitutive equations is of prior importance 
\cite{silverman} and 
 was itself of phenomenological nature until the link between the
 optical activity as well as the magnetic-dipole and electric-octupole responses 
of the crystal medium
 was consistently formalised\cite{raab1,raab2}. This theory leads to
 constitutive equations different from those previously used.  
 To the author's knowledge, the constitutive equations of Ref. \cite{raab1}
 have not yet been used to calculate
 the transmission matrix of an optical active uniaxial platelet. They were only 
applied to describe 
  the reflection of a plane wave by an interface between an 
 isotropic achiral and a uniaxial chiral media \cite{raab-reflection}, but
 for an optical axis parallel or perpendicular to the plane of incidence.
  It is the purpose of this article to provide this transmission matrix in
 a simple and usable form. We further
 restrict ourselves to two geometric configurations
 often encountered experimentally, namely
  the optical axis parallel and perpendicular to the plane of interface.

 Recent experimental
 results have shown that the contribution of the optical activity to the 
transmission
 matrix of a quarter wave plate can reach the percent level \cite{chou}. As an
 application of our transmission matrix formula, the origin
 of this effect is investigated and the case of half wave plates is also 
considered. 
 Our formula being valid under oblique incidence, numerical calculations of the 
variations of the
 platelet transmission matrix as a function
 of the angle of incidence are presented.

 In section \ref{theory}
 we describe the method. In section \ref{deux}, we derive the
 explicit expressions of the platelet transmission matrix for oblique and
 normal incidences. The contribution of the optical activity to 
 the transmission matrix is studied numerically for various quartz
 platelet thicknesses in section \ref{trois}.

\section{Formalism}\label{theory}
A monochromatic plane wave of wavelength $\lambda$ impinging a 
 uniaxial optically active crystal slab of thickness $\ell$
 is considered. The crystal is assumed to be non-absorbing, non-magnetic
 and surrounded by an isotropic achiral ambient medium.

\begin{figure}[t]
\centering
\includegraphics{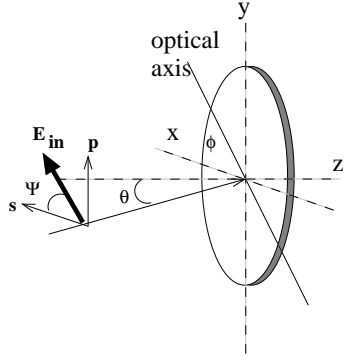}
\caption{Schematic view of an optically active platelet with its optical axis in the plane of interface. The angles and axes used in the calculations are also drawn.}
\label{geom}
\end{figure}

 A direct axis system $x,y,z$
 is defined such that the $z$ axis is perpendicular to the plane
 of interface. The origin of the $z$ axis ({\it i.e.} $z=0$) is fixed on the 
first plane
 of interface so that $z=\ell$ corresponds to the second plane of interface (see Fig. \ref{geom}).
The unit vector basis attached to the system axes
 is denoted $\{\xhat,\yhat,\zhat\}$.Without loss of generality, the plane of incidence is
 taken to be $yz$. To determine the transmission and reflection matrices
 of the interfaces between the ambient medium and the crystal faces,
 we follow the method of Ref. \cite{yeh2x2}.
\clearpage
\vspace{2mm}
 Near the first interface $z=0$, the electric and magnetic field vectors read:
\begin{align}
\hbox{Incident: }
\Ev=&
\biggl[A_s\shat+A_p\phat\biggr]\exp(i[\omega t-\kvec\cdot\er])
\label{incident}\\
\Hv=&\frac{1}{\mu_0\omega}\kvec\times
\biggl[A_s\shat+A_p\phat\biggr]\exp(i[\omega t-\kvec\cdot\er])\\
\hbox{ Reflected: }
\Ev=&
\biggl[B_s\shat+B_p\phat'\biggr]
\exp(i[\omega t-\kvec'\cdot\er])\\
\Hv=&\frac{1}{\mu_0\omega}\kvec'\times
\biggl[B_s\shat+B_p\phat'\biggr]\exp(i[\omega t-\kvec'\cdot\er])\\
\hbox{Refracted: }
\Ev=&\biggl[C_o\ovec_+\exp(-i\kveco_+\cdot\er)+
C_e\evec_+\exp(-i\kvece_+\cdot\er)\biggr]
\exp(i\omega t)\label{incidentE}\\
\Hv=&\frac{1}{\mu_0\omega}
\biggl[C_o(\kveco_+\times\ovec_+-i\mu_0\omega
{\mathcal T}\ovec_+)\exp(-i\kveco_+\cdot\er)+\nonumber\\
&C_e(\kvece_+\times\evec_+-i\mu_0\omega
{\mathcal T}\evec_+)\exp(-i\kvece_+\cdot\er)\biggr]
\exp(i\omega t)\label{incidentH}
\end{align}
and near the second interface $z=\ell$:
\begin{align}
\hbox{Reflected: }
\Ev=&\biggl[C'_o\ovec_-\exp(-i\kveco_-\cdot\er)+
C'_e\evec_-\exp(-i\kvece_-\cdot\er)\biggr]
\exp(i\omega t)\\
\Hv=&\frac{1}{\mu_0\omega}
\biggl[C'_o(\kveco_-\times\ovec_--i\mu_0\omega
{\mathcal T}\ovec_-)\exp(-i\kveco_-\cdot\er)+\nonumber\\
&C'_e(\kvece_-\times\evec_--i\mu_0\omega
{\mathcal T}\evec_-)\exp(-i\kvece_-\cdot\er)\biggr]
\exp(i\omega t)\\
\hbox{Refracted: }
\Ev=&
\biggl[A'_s\shat+A'_p\phat\biggr]\exp(i[\omega t-\kvec\cdot\er])
\label{transmited}\\
\Hv=&\frac{1}{\mu_0\omega}\kvec\times
\biggl[A'_s\shat+A'_p\phat\biggr]\exp(i[\omega t-\kvec\cdot\er])
\end{align} 
with the incident field vectors given by
 Eqs. (\ref{incidentE},\ref{incidentH}).
\newpage
In these expressions,  
 $\kvec$ and $\kvec'$ are the wave vectors in the ambient medium;
 the three vectors $\shat$, $\phat$ and $\kvec$ form a 
 direct basis, they are given by
 $\shat=\phat\times\kvec/|\phat\times\kvec|$ and
 $\shat=\phat'\times\kvec'/|\phat'\times\kvec'|$ 
 with $\shat$ perpendicular to the plane of incidence 
({\it i.e.} $\shat=\xhat$); ${\mathcal T}$ is a second-rank tensor 
 coming from the constitutive equations \cite{raab1},
 its components are related to those of the gyration
 tensor coefficients $g_{ij}$ \cite{ny}; the electric vectors $\ovec_\pm$
 and $\evec_\pm$ are the solutions of the wave equation inside the crystal; 
 the wave vectors inside the medium, 
 $\kveco_\pm$ and $\kvece_\pm$, are given by the Snell's law and by the
 condition of existence of solutions of the wave equation.

The wave equation inside the crystal is given by \cite{raab1}
\begin{equation}\label{wave}
\sum_\beta 
\biggl(k_\alpha k_\beta -\delta_{\alpha\beta} 
k^2+\mu_0\omega^2\epsilon_{\alpha\beta}
 +i\mu_0\omega \sum_\gamma k_\gamma {\mathcal A}_{\alpha\beta\gamma}\biggr)
E_\beta=0
\end{equation}
where $\alpha,\beta$ and $\gamma$ stand for $x$ or $y$ or $z$
 and where the magnetic-dipole and electric-quadrupole response of the medium
 is embodied in the third-rank tensor ${\mathcal A}$
 whose components are related to those of the gyration 
tensor\cite{raab1,raab-reflection}.
 In Eq. (\ref{wave}), $k_\alpha$ stands for
 the components of one of the four wave vectors $\kveco_\pm$ and $\kvece_\pm$;
 $E_\beta$ stands for one of the components of the four
 electric vectors $\ovec_\pm$
 and $\evec_\pm$; $\epsilon_{\alpha\beta}$ are the elements of the dielectric 
tensor.

The wave amplitudes are related by matrix relations. The refraction by the
first ({\it i.e.} $z=0$) and second ({\it i.e.} $z=\ell$)
 interfaces are described
by \cite{yeh2x2}:
\begin{equation}
\begin{pmatrix}
C_o \\
C_e 
\end{pmatrix}
=T
\begin{pmatrix}
A_s \\
A_p 
\end{pmatrix}\,\hbox{ and }
\begin{pmatrix}
A'_s \\
A'_p 
\end{pmatrix}
=T'P_+
\begin{pmatrix}
C_o \\
C_e 
\end{pmatrix}
\end{equation}
respectively. As for the reflections by the first and second 
 crystal-ambient medium interfaces (inside the crystal), one gets
\begin{equation}
\begin{pmatrix}
C_o \\
C_e 
\end{pmatrix}
=R
\begin{pmatrix}
C'_o \\
C'_e 
\end{pmatrix}\,\hbox{ and }
\begin{pmatrix}
C'_o \\
C'_e 
\end{pmatrix}
=P_-^{-1}RP_+
\begin{pmatrix}
C_o \\
C_e 
\end{pmatrix}
\end{equation}
respectively. In the previous equations, the following \tbyt  phase matrices
 were introduced:
\begin{equation}
P_\pm=
\begin{pmatrix}
\exp(-i\varphi_{o_\pm})&0 \\
0&\exp(-i\varphi_{e_\pm}) \\ 
\end{pmatrix},\,
\end{equation}
with $\varphi_{o_\pm}=\ell(\kveco_\pm\cdot\zhat)$ and
 $\varphi_{e_\pm}=\ell(\kvece_\pm\cdot\zhat)$.

The \tbyt matrices $T$, $T'$ and $R$ are obtained from the 
 continuity conditions of the projections of the electric and magnetic
 field vectors in the planes $z=0$ and $z=\ell$. Once known, these
 matrices are used to compute the platelet transmission matrix $M$ such that 
$\Evo=M\Evi$ where
 $\Evi$ and $\Evo$ are the electric field vectors before and after the crystal
 slab respectively. Note that $M$, $\Evi$ and $\Evo$ are thus represented in the
 $\{\shat,\phat,\khat\}$ basis, so that $M$ reduces to a \tbyt  matrix
 and $\Evi$ and $\Evo$ to two-dimension vectors when plane waves are considered.

 Taking into account the multiple reflections inside the
 crystal, one gets:
\begin{equation}\label{jones}
M=T'P_+T+T'P_+RP_-^{-1}RP_+T+\cdots=T'P_+\biggl[\un-RP_-^{-1}RP_+
\biggr]^{-1}T
\end{equation} 
where we have used $\sum_{i=0}^{\infty}X^i=[\un-X]^{-1}$ and where $\un$ is the 
\tbyt  
identity matrix.

 We shall now choose a crystal symmetry
 class group in order to specify the two tensors ${\mathcal A}$ and ${\mathcal 
T}$.

\section{Transmission matrix of a platelet with the optical axis parallel 
 and perpendicular to the plane of interface}\label{deux}
Quartz crystal is widely used in crystallography to test the 
 reliability of experimental setups \cite{hernandez,chou} and to manufacture 
retardation plates.
 We thus restrict ourselves to the non-centrosymmetric
crystals belonging to the symmetry classes 32, 422 and 622
 \cite{ny,briss} for which the tensors
 ${\mathcal A}$ and ${\mathcal T}$ have similar expressions 
 \cite{raab-reflection}. In the crystallographic reference frame,
 {\it i.e} for the crystal optical axis aligned along the $z$ axis,
 the non-vanishing components of these tensors 
 are \cite{raab-reflection,briss}: 
\begin{align}
&\tilde{\mathcal T}_{xx}=\tilde{\mathcal T}_{yy}=g_{33}/2,\,
\tilde{\mathcal T}_{zz}=g_{11}-g_{33}/2\\
&\tilde{\mathcal A}_{xyz}=-\tilde{\mathcal A}_{yxz}=g_{33},\,\tilde{\mathcal 
A}_{yzx}=-
\tilde{\mathcal A}_{zyx}
=g_{11},\,
\tilde{\mathcal A}_{zxy}=-\tilde{\mathcal A}_{xzy}=g_{11}
\end{align}
with $\tilde{\mathcal T}_{lm}=\mu_0c{\mathcal T}_{lm}$ and
 $\tilde{\mathcal A}_{lmn}=\mu_0c{\mathcal A}_{lmn}$ and where 
$g_{11}$ and $g_{33}$ are the two independent coefficients of
 the gyration tensor\cite{ny}. 
 Note that for another orientation of the optical axis,
 standard tensor transformations \cite{ny,briss} are used to determine
 the representations of $\tilde{\mathcal T}$ and $\tilde{\mathcal A}$.
 
We shall now consider two geometrical configurations often
 encountered experimentally: a platelet with its optical axis parallel 
 and perpendicular to the plane of interface. The two cases are treated 
separately
 because different approximations are made during the calculations.

\subsection{Optical axis parallel to the plane of interface}\label{oblique}

The wave equation (\ref{wave}) can be written in a matrix form
\begin{equation}\label{wavemat}
({\mathcal E}+\delta{\mathcal E})\Ev=0
\end{equation}
where the contribution of the optical activity is contained in the second term 
$\delta{\mathcal E}$ and 
 where $\Ev=(E_x,E_y,E_z)^T$.
Writing the wave vector $\kvec=(\omega/c)(0,\betat,\kzt)^T$, with 
$\betat=n_a\sin\theta$ and where
 $\theta$ is the angle of incidence and $n_a$ the optical index of the ambient 
medium, we obtain:
\[
{\mathcal E}=
\begin{pmatrix}
n_e^2\cos^2\phi + n_o^2\sin^2\phi-\kzt^2-\betat^2 & 
\Delta\sin\phi\cos\phi & 0\\
\Delta\sin\phi\cos\phi&
 n_o^2\cos^2\phi + n_e^2\sin^2\phi-\kzt^2 & \betat\kzt\\
0 & \betat\kzt & n_o^2-\betat^2
\end{pmatrix}
\]
and 
\[
\delta{\mathcal E}=
\begin{pmatrix}
0 & ig_{11}\kzt & -i\betat(g_{11}\cos^2\phi+g_{33}\sin^2\phi)\\
-ig_{11}\kzt & 0 & i\betat\sin\phi\cos\phi(g_{33}-g_{11})\\
i\betat(g_{11}\cos^2\phi+g_{33}\sin^2\phi) &-i\betat\sin\phi\cos\phi(g_{33}-
g_{11}) & 0
\end{pmatrix}
\]
with $\Delta=n_e^2-n_o^2$ and where
 $n_o=\sqrt{\epsilon_o/\epsilon_0}$ and $n_e=\sqrt{\epsilon_e/\epsilon_0}$
 are the ordinary and extraordinary optical indices respectively. In the 
previous expressions,
 $\phi$ is the azimuth angle describing the orientation of the optical axis 
 (note that $\phi=0$ corresponds to the optical axis perpendicular to 
 the plane of incidence). 
 
Since optical activity is induced by the electric-quadrupole and magnetic-dipole
 responses of the crystal medium\cite{raab1},
 we assume that $|g_{ij}|\ll |n_e-n_o|$. The condition of existence of a 
solution
 for Eq. (\ref{wavemat}) reads $det({\mathcal E}+\delta{\mathcal E})=0$. This 
condition gives us
 the four possible values for $\kzt$:
${\tilde k}_{oz_\pm}=\pm({\tilde k}_{oz}^{(0)}+\delta{\tilde k}_{oz})$ and
${\tilde k}_{ez_\pm}=\pm({\tilde k}_{ez}^{(0)}+\delta{\tilde k}_{ez})$ which 
correspond to the ordinary and
 extraordinary waves respectively (the sign $\pm$ refers to the direction of 
propagation). The zero 
 order terms ${\tilde k}_{oz}^{(0)}$ and ${\tilde k}_{ez}^{(0)}$ are given by 
the solution of
 $det({\mathcal E})=0$:
\begin{align}
{\tilde k}_{oz}^{(0)}&=(n_o^2-\betat^2)^{1/2},\\
{\tilde k}_{ez}^{(0)}&=\biggl(n_e^2-
\betat^2(\cos^2\phi+\frac{n_e^2}{n_o^2}\sin^2\phi)\biggr)^{1/2},
\end{align}
and the first non-vanishing contributions in $g_{ij}$ read as
\begin{align}
\delta{\tilde k}_{oz}&=\frac{1}{4n_o^2{\tilde k}_{oz}^{(0)}}(g_{11}^2n_o^2-
\betat^2\sin^2\phi
(g_{33}-g_{11})^2-\delta k)\label{ko}\\
\delta{\tilde k}_{ez}&=\frac{1}{4n_o^2{\tilde k}_{ez}^{(0)}}(g_{11}^2n_o^2-
\betat^2\sin^2\phi
(g_{33}-g_{11})^2+\delta k)\label{ke}
\end{align}
with
\begin{eqnarray}
\delta k=\frac{1}{(n_o^2-
\betat^2\sin^2\phi)\Delta}\biggl[n_o^4g_{11}^2(n_e^2+n_o^2)\nonumber\\
+\betat^2 n_o^2\sin^2\phi [2g_{11}(g_{33}-
g_{11})(n_e^2+n_o^2)+g_{33}^2\Delta]\nonumber\\
+\betat^4\sin^4\phi(g_{33}-g_{11})^2(n_e^2+n_o^2)\biggr].
\end{eqnarray}
As expected \cite{raab1}, Eqs. (\ref{ko}) and (\ref{ke}) are of second order in 
$g_{ij}$.

Accordingly, the ordinary and extraordinary electric field vectors are 
decomposed as follows:
 $\ovec_{\pm}=\ovec^{(0)}_{\pm}+\delta \ovec_{\pm}$ and  
$\evec_{\pm}=\evec^{(0)}_{\pm}+\delta \evec_{\pm}$.
 The zero order terms $\ovec^{(0)}_{\pm}$ and $\evec^{(0)}_{\pm}$ are solutions 
of 
  ${\mathcal E}\ovec^{(0)}_{\pm}=0$ and ${\mathcal E}\evec^{(0)}_{\pm}=0$:  
\begin{align}
\ovec^{(0)}_{\pm}&=N_o({\tilde k}_{oz}^{(0)}\sin\phi,-{\tilde 
k}_{oz}^{(0)}\cos\phi,\pm\betat\cos\phi)^T,\\
\evec^{(0)}_{\pm}&=N_e(n_o^2\cos\phi,({\tilde 
k}_{oz}^{(0)})^2\sin\phi,\mp{\tilde k}_{ez}^{(0)}\sin\phi)^T.
\end{align}
where $N_o$ and $N_e$ are the normalisation factors such that 
$|\ovec^{(0)}_{\pm}|=1$ and 
 $|\evec^{(0)}_{\pm}|=1$.
 $\delta \ovec_{\pm}$ and $\delta \evec_{\pm}$, the first order contributions in 
$g_{ij}/\Delta$,
 are the solutions of 
 ${\mathcal E}\delta \ovec_{\pm}=-\delta{\mathcal E}\ovec^{(0)}_{\pm}$ and
 ${\mathcal E}\delta \evec_{\pm}=-\delta{\mathcal E}\evec^{(0)}_{\pm}$. Although 
${\tilde k}_{oz\pm}$ 
 and ${\tilde k}_{ez\pm}$ are of second order in
 $g_{ij}$, implying that
 $det({\mathcal E})=0$ at first order, it turns out that the conditions 
 of existence of solutions for these equations are fulfilled. We obtain
\begin{align}
\delta \ovec_{\pm}&=
\frac{\pm iN_o}{\Delta}
\begin{pmatrix}
\cos\phi[g_{11}n_o^2+\betat^2\sin^2\phi(g_{33}-g_{11})]\\
\sin\phi[g_{11}n_o^2+\betat^2\sin^2\phi(g_{33}-g_{11})]\\
-\betat\sin\phi[g_{11}n_o^2+g_{33}\Delta+\betat^2\sin^2\phi(g_{33}-
g_{11})]/(n_o^2-\betat^2)
\end{pmatrix}
\label{dovecprime}\\
\delta \evec_{\pm}&=
\frac{\pm iN_e[g_{11}n_o^2+\betat^2\sin^2\phi(g_{33}-g_{11})]}{\Delta}
\begin{pmatrix}
{\tilde k}_{ez}^{(0)}\sin\phi n_o^2/(n_o^2-\betat^2)\\
-{\tilde k}_{ez}^{(0)}\cos\phi\\
\betat\cos\phi[({\tilde k}_{ez}^{(0)})^2-\Delta]/(n_o^2-\betat^2)^2
\end{pmatrix}
\label{devecprime}
\end{align}

Once the ordinary and extraordinary wave and electric vectors are known, the 
matrices ${\mathcal R}$, ${\mathcal T}$ 
 and ${\mathcal T}'$ are determined by supplying the boundary conditions in the 
planes $z=0$ and $z=\ell$. 
 For an oblique incidence, these calculations are more simply performed 
numerically. Compact analytical 
 expressions are obtained for the special case of normal incidence and will be 
given bellow.

To define a platelet transmission matrix at first order
 in $g_{ij}/\Delta$, we introduce the three matrices $R_0$, $T_0$ and $T'_0$
 obtained by setting $g_{11}=g_{33}=0$ to zero in the expressions of $R$, $T$ 
and $T$ respectively.
 We then define the matrices $\delta R$, and $\delta T$ and $\delta T'$ such
 that $R=R_0+\delta R$, $T=T_0+\delta T$ and $T'=T'_0+\delta T$. Having thus
 isolated the zero order from the first order terms in 
 $g_{ij}/\Delta$, we define the platelet transmission matrix $M=M_0+\delta M$
 and thereby
\begin{equation}\label{deltaE}
 \Evo=\Evoo+\delta\Evo \hbox{ such }\Evoo=M_0\Evi,\,
 \delta\Evo=\delta M \Evi. 
\end{equation}
 Proceeding as in Eq. (\ref{jones}), we obtain: 
\begin{align}\label{deltaM}
\delta M=&\delta T'(T'_0)^{-1}M_0+M_0T_0^{-1}\delta T+\nonumber\\
&T'_0\biggl[\un-P_+R_0P_-^{-1}R_0\biggr]^{-1}
P_+R_0P_-^{-1}\delta R
\biggl[\un-P_+R_0P_-^{-1}R_0\biggr]^{-1}
P_+T_0+\nonumber\\ 
&T'_0\biggl[\un-P_+R_0P_-^{-1}R_0\biggr]^{-1}
P_+\delta R\biggl[\un-P_-^{-1}R_0P_+R_0\biggr]^{-1}P_-^{-1}R_0P_+T_0
\end{align}
 and    
\begin{equation}\label{M0}
M_0=T'_0P_+
\biggl[\un-R_0P_-^{-1}R_0P_+\biggr]^{-1}T_0 .
\end{equation} 
 Note that $M_0$ only depends on
 the optical activity through the phase matrices $P_+$ and $P_-^{-1}$.

Under normal incidence, one gets
 ${\tilde k}_{oz_\pm}=\pm n_o[1-g^2_{11}/(2\Delta)]$,
 ${\tilde k}_{ez_\pm}=\pm n_e[1+g^2_{11}/(2\Delta)]$ to the first order in 
$g_{11}/\sqrt{\Delta}$
and the corresponding electric vectors 
\begin{align} 
\ovec_\pm&=(\sin\phi\pm \frac{in_og_{11}\cos\phi}{\Delta},
-\cos\phi\pm \frac{in_og_{11}\sin\phi}{\Delta},0)^T,\\ 
\evec_\pm&=(\cos\phi\pm \frac{in_eg_{11}\sin\phi}{\Delta},
\sin\phi\mp \frac{in_eg_{11}\cos\phi}{\Delta},0)^T.
\end{align}

The reflection and transmission matrices are:
\begin{align}
R&=\begin{pmatrix}
\frac{n_o-n_a}{n_o+n_a} & 
in_e\frac{2g_{11}(n_o^2-n_a^2)+g_{33}\Delta}{(n_a+n_o)(n_a+n_e)\Delta} \\
in_o\frac{2g_{11}(n_o^2-n_a^2)+g_{33}\Delta}{(n_a+n_o)(n_a+n_e)\Delta} &
\frac{n_e-n_a}{n_e+n_a}
\end{pmatrix}
\\
T&=\begin{pmatrix}
-in_a\frac{2g_{11}(n_en_a+n_o^2)+g_{33}\Delta}{(n_a+n_o)(n_a+n_e)\Delta}
&\frac{-2n_a}{n_o+n_a}  
 \\
\frac{2n_a}{n_e+n_a}&
in_a\frac{2g_{11}n_o(n_a+n_o)+g_{33}\Delta}{(n_a+n_o)(n_a+n_e)\Delta} 
\end{pmatrix}
{\mathcal R}(-\phi)
\label{T}\\
T'&=
{\mathcal R}(\phi)
\begin{pmatrix}
in_o\frac{2g_{11}(n_en_a+n_o^2)+g_{33}\Delta}{(n_a+n_o)(n_a+n_e)\Delta}
&\frac{2n_e}{n_e+n_a}  
 \\
\frac{-2n_o}{n_o+n_a}&
-in_e\frac{2g_{11}n_o(n_a+n_o)+g_{33}\Delta}{(n_a+n_o)(n_a+n_e)\Delta} 
\end{pmatrix}
\end{align}
where ${\mathcal R}(\phi)$ is the $\tbyt$ matrix representing a rotation
 of an angle $\phi$ around the $z$ axis.

In these expressions, terms
 in $g_{11}$ are enhanced, unlike those in $g_{33}$, by a factor
 $1/(n_e-n_o)\gg 1$. This is a major difference with the case of an optical axis 
perpendicular
 to the plane of interface (see next section). From
 Eqs. (\ref{dovecprime}-\ref{devecprime}) 
  one also sees that the contribution of $g_{33}$ to the transmission matrix
 becomes noticeable only under oblique incidence. 

 Let us mention that Eq. (\ref{T}) agrees with the result of
 Ref. \cite{silverman} for an isotropic chirality, {\it i.e.} $g_{11}=g_{33}$. 
It is
 however in disagreement with the expression of Ref. \cite{miteva}
 obtained using phenomenological constitutive equations. The expression of our
 platelet transmission matrix
 therefore disagrees with the one of Refs. \cite{miteva2,miteva3}. 

The zero and first order platelet transmission matrices are given by Eqs. 
(\ref{M0}) and (\ref{deltaM}) 
 respectively. For Eq. (\ref{M0}), we obtain the following simple 
expression\cite{zeller}: 
\begin{equation}\label{jones0}
M_0=
{\mathcal R}(\phi)
\begin{pmatrix}
\frac{4n_on_a\exp(-i\varphi_o)}{(n_a+n_o)^2-(n_a-n_o)^2\exp(-2i\varphi_o)} &0\\
0&\frac{4n_en_a\exp(-i\varphi_e)}{(n_a+n_e)^2-(n_a-n_e)^2\exp(-2i\varphi_e)}
\end{pmatrix}
{\mathcal R}(-\phi)
\end{equation}
with $\varphi_o= (\omega/c)\ell {\tilde k}_{oz_+}$ and
 $\varphi_e= (\omega/c)\ell {\tilde k}_{ez_+}$.

\subsection{Optical axis perpendicular to the plane of interface}\label{perp}
When the optical axis is parallel to the $z$ axis, one
 can write Eq. (\ref{wave}) as follows:
\[
\begin{pmatrix}
n_o^2-{\tilde k}_z^2 -{\tilde\beta}^2& ig_{33}{\tilde k}_z &
 -ig_{11}{\tilde\beta}\\
-ig_{33}{\tilde k}_z & n_o^2-{\tilde k}^2_z & {\tilde\beta }{\tilde k}_z \\
ig_{11}{\tilde\beta} & {\tilde\beta }{\tilde k}_z & n_e^2-{\tilde\beta}^2
\end{pmatrix}
\begin{pmatrix}
E_x\\
E_y\\
E_z
\end{pmatrix}
=0.
\] 
To first order in $g_{ij}$, the four solutions for $k_z$ are
 given by ${\tilde k}_{oz_\pm}=\pm(a-b)^{1/2}$ and
 ${\tilde k}_{ez_\pm}=\pm(a+b)^{1/2}$ with
\begin{align}
a&=n_o^2-\frac{{\tilde\beta}^2}{2n_e^2}(n_o^2+n_e^2)\\
b&=\biggl(n_o^2g_{33}^2+\frac{n_o^2g_{33}{\tilde\beta}^2}{2n_e^2}
(4g_{11}-3g_{33})+\frac{{\tilde\beta}^4}{4n_e^4}
[2(n_o^2+n_e^2)(g_{33}-g_{11})^2+(n_e^2-n_o^2)^2]
\biggr)^{1/2}
\end{align}
and the corresponding electric vectors read as \cite{yeh4x4}
\[
\vhat=
N\begin{pmatrix}
n_e^2(n_o^2-{\tilde k}_z^2)-n_o^2{\tilde\beta}^2\\
i{\tilde k}_zg_{33}n_e^2+i{\tilde\beta}^2{\tilde k}_z(g_{33}-g_{11})\\
-i{\tilde\beta}[{\tilde k}_z^2(g_{33}-g_{11})+g_{11}n_o^2]
\end{pmatrix}
\] 
where $\vhat\equiv\ovec_\pm,\,\evec_\pm$ for
 ${\tilde k}_z\equiv{\tilde k}_{oz_\pm},\,{\tilde k}_{ez_\pm}$
 respectively and where $N$ is the normalisation factor such that $|\vhat|=1$.

As in the previous case, the transmission matrix is more simply computed
 numerically. Under normal incidence,
 one gets ${\tilde k}_{oz_\pm}=\pm(n_o-g_{33}/2)$ and
 ${\tilde k}_{ez_\pm}=\pm(n_o+g_{33}/2)$ to the first order in $g_{33}$. 
The corresponding electric vectors are the two circular polarisation states 
 that we write $\ovec_\pm^T=(\mp i,1,0)/\sqrt{2}$ and 
$\evec_\pm^T=(\pm i,1,0)/\sqrt{2}$.
For the transmission and reflection matrices, one gets:
\[
T=\frac{\sqrt{2}n_a}{n_a+n_o}
\begin{pmatrix}
i & 1\\
-i & 1
\end{pmatrix},\,
T'=\frac{n_o\sqrt{2}}{n_a+n_o}
\begin{pmatrix}
-i & i\\
1 & 1
\end{pmatrix},\,
R=\frac{n_o-n_a}{n_o+n_a}
\begin{pmatrix}
0 & 1\\
1 & 0
\end{pmatrix}
.
\]
Using Eq. (\ref{jones}), the following platelet transmission matrix is obtained: 
\begin{equation}\label{matp}
M=\frac{2n_on_a[\exp(-i\varphi_o)+\exp(-i\varphi_e)]}
{(n_a+n_o)^2-(n_a-n_o)^2\exp(-i[\varphi_o+\varphi_e])}
\begin{pmatrix}
1 & -\tan((\varphi_{e}-\varphi_o)/2)\\
 \tan((\varphi_{e}-\varphi_o)/2) & 1
\end{pmatrix}
\end{equation}
with $\varphi_o=\ell k_{oz_+}$ and $\varphi_e=\ell k_{ez_+}$.
 This expression agrees
 with the result obtained using a \fbyf matrix method \cite{lautre}.
 
Note that the interface matrices of eq. (\ref{matp}) do not depend
 explicitly of the coefficients $g_{11}$ and $g_{33}$. Thereby, the variations
 of $M$ with these coefficients come only from the phases  $\varphi_o$
 and $\varphi_e$. Note as well that the coefficient $g_{11}$ contributes only 
under
 oblique incidence.
\vspace{-5mm}
\section{Numerical results}\label{trois}
 \vspace{-5mm}
 To estimate the effects of the gyration
 coefficients on the platelet transmission,
 an experience closed to the HAUP\cite{hernandez} is considered: a
 linearly polarised He-Ne laser beam
  crosses a quartz platelet and then a supposed perfect analyser.
 The incident electric
 vector reads $\Evi=(\cos\psi,\sin\psi)^T$ in the $\{\shat,\phat\}$ basis
 ({\it i.e.} in the basis $\{\xhat,\yhat\}$ under normal incidence, see Fig. 
\ref{geom}). 
 After the analyser, whose eigen axis corresponds
 to, say, the $x$ axis, the beam intensity is  measured as a function of $\psi$
 and $\phi$. 

For our wavelength, $\lambda=0.6328\mu$m, the optical indices are\cite{sosman}
 $n_o=1.542637$ and $n_e=1.551646$ and the 
 gyration coefficients $g_{11}=-5.9\cdot 10^{-5}$ and $g_{33}=10.1\cdot 10^{-
5}$\cite{hernandez}.

To first order in $g_{ij}/\Delta$,
 the beam intensity after the analyser is given by:
\begin{equation}\label{Ix-numeric}
I_x=|\Evoo\cdot\xhat|^2+[(\Evoo\cdot\xhat)
(\delta\Evo\cdot\xhat)^*+
(\Evoo\cdot\xhat)^*(\delta\Evo\cdot\xhat)]
\end{equation}
 where the field vectors are defined in Eq.
 (\ref{deltaE}).

 We first consider the normal incidence and a crystal
 optical axis parallel to the plane of interface.
 As for the plate thickness, two values are chosen:
 $\ell_1=87.80\mu$m and $\ell_2=1000.93\mu$m ({\it i.e.} a first and a fourteen 
order
 quarter wave plates). 

\begin{figure*}[h]
\centering
\includegraphics{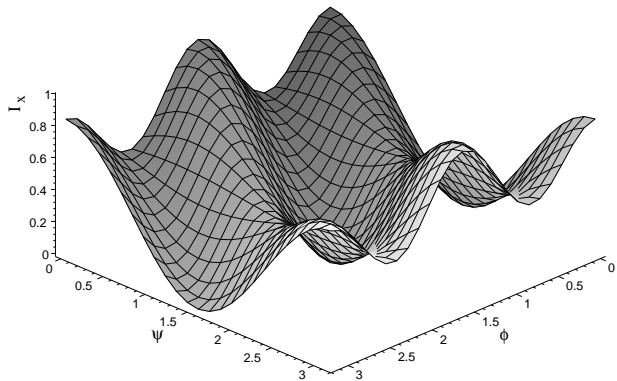}
\caption{$I_x$, for a normal incidence angle and
 for the plate thickness $\ell_1$, as a function of the angles $\phi$ and $\psi$ 
(in radian).
 The optical axis is parallel
 to the plane of interface.}
    \label{I1}
\end{figure*}

\begin{figure*}[h]
\centering
\includegraphics{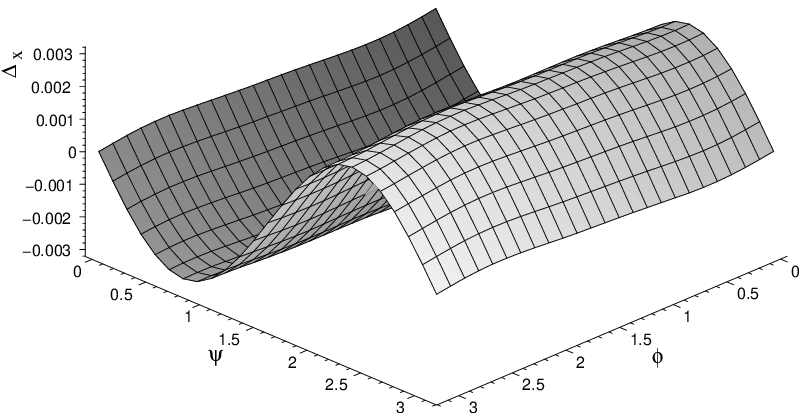}
\caption{ $\Delta_x$, for a normal incidence angle and
 for the plate thickness $\ell_1$, as a function of the angles $\phi$ and $\psi$ 
(in radian).
 The optical axis is parallel
 to the plane of interface.}
    \label{DI1_x}
\end{figure*}

 The intensity $I_x$
 is shown as a function of $\psi$ and $\phi$ in Fig. \ref{I1}
 (it is similar for the two plate thicknesses $\ell_1$ and $\ell_2$).
 Defining $I_{0x}=|\xhat\cdot\Evoo|^2=|\xhat\cdot M_0\Evi|^2$, the difference
$\Delta_{x}=I_x-I_{0x}$ is shown
 as a function of $\phi$ and $\psi$ for the plate thickness $\ell_1$
 in Fig. \ref{DI1_x}. The effect due to the medium gyrotropy is surprisingly
 high for $\psi\approx\pi/4$ and the same behaviour is indeed
 observed for the second plate thickness $\ell_2$. In the expression
 of $I_{0x}$, the contribution of the 
 optical activity comes only from the phases
 $\varphi_o$ and $\varphi_e$.

\begin{figure}[h]
\centering
\includegraphics{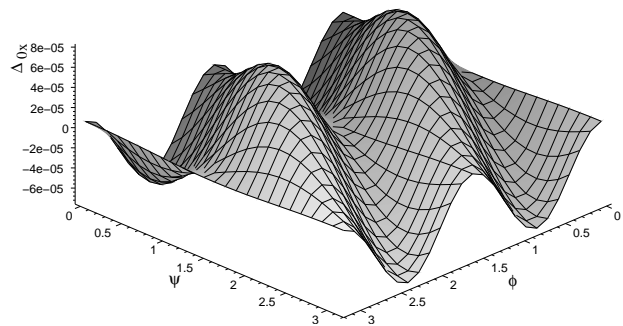}
\caption{$\Delta_{0x}$, for a normal incidence angle and
 for the plate thickness $\ell_1$, as a function of the angles $\phi$ and $\psi$ 
(in radian).
 The optical axis is parallel
 to the plane of interface.}
    \label{Itilde1_x}
\end{figure}

To distinguish the chiral contribution coming from the phases
 $\varphi_o$ and $\varphi_e$ from the one coming from the interface matrices,
  we define 
${\tilde I}_{0x}=|\xhat\cdot{\tilde M}_0\Evi|^2$ where ${\tilde M}_0$ is
 given by Eq. \ref{jones0}, but fixing $\varphi_o=2\pi n_o/\lambda$ and
 $\varphi_e=2\pi n_e/\lambda$
 ({\it i.e.} by neglecting the optical activity). 
 The difference $ \Delta_{0x}=I_{0x}-{\tilde I}_{0x}$ is shown in Fig. \ref{Itilde1_x} as a function of $\psi$ and $\phi$ for the plate thicknesses $\ell_1$. These figures indicate that the large variations observed in Fig. \ref{DI1_x} come
 from the chiral dependence of the interface matrices.
 For the plate thickness $\ell_2$, the shape of $\Delta_{0x}$ is identical
 to Fig. \ref{Itilde1_x} but scaled by a factor of ten.  
 As expected, the chiral dependence induced by the phase shift alone becomes 
only sizeable for
 thick plates.

\begin{figure}[h]
\centering
\includegraphics{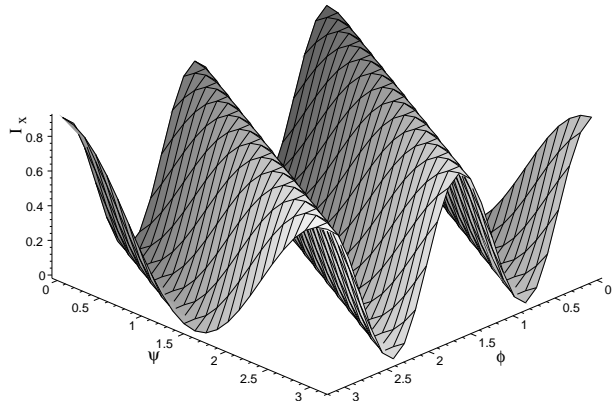}
\caption{$I_x$, for a normal incidence angle and for the plate thickness
$\ell_3$, as a function of the angles $\phi$ and $\psi$(in radian). 
The optical axis is parallel to the plane of interface.}
\label{I2}
\end{figure}
\begin{figure}[h]
\centering
\includegraphics{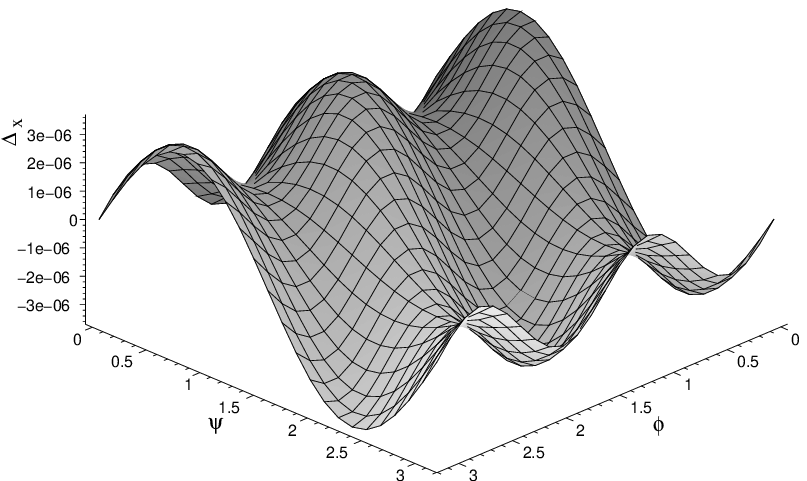}
\caption{$\Delta_x$, for a normal incidence angle and
 for the plate thickness $\ell_3$, as a function of the angles $\phi$ and $\psi$ 
(in radian).
 The optical axis is parallel
 to the plane of interface.}
    \label{DI2_x}
\end{figure}
\begin{figure}[h]
\centering
\includegraphics{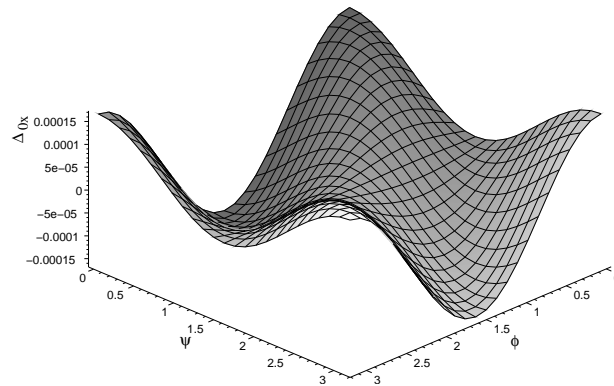}
\caption{$\Delta_{0x}$, for a normal incidence angle and
 for the plate thickness $\ell_3$, as a function of the angles $\phi$ and $\psi$ 
(in radian).
 The optical axis is parallel
 to the plane of interface.}
    \label{Itilde2_x}
\end{figure}

The previous discussion holds for quarter wave plates. Let us consider another 
plate thickness
$\ell_3=1018.49\mu$m, {\it i.e.} a fourteen order half wave plate. $I_x$, 
$\Delta_{x}$ and $ \Delta_{0x}$ 
 are shown as a function of $\phi$ and $\psi$ in Figs. \ref{I2}, \ref{DI2_x} and  
\ref{Itilde2_x}
 respectively.
 The effect induced by the optical activity is here reduced by two orders of 
magnitude 
 with respect to a quarter wave plate and is dominated by the 
 variation of the phases $\varphi_o$ and $\varphi_e$ with $g_{11}$.
 For a first order half wave plate ($\ell_4=105.36\mu$m), the effect on 
$\Delta_{0x}$ is reduced by
 an order of magnitude whereas $\Delta_{x}$ is essentially unchanged.

\begin{figure}[h]
\centering
\includegraphics{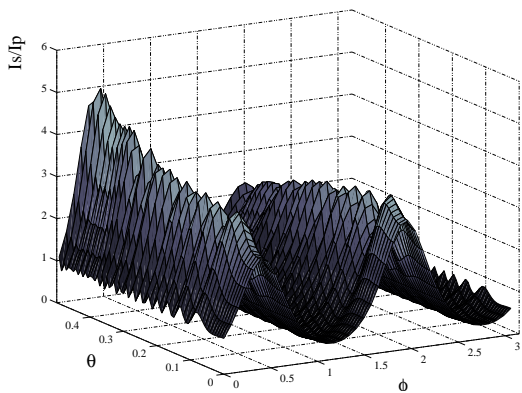}
\caption{$r$ as a function of the angles $\theta$ and $\phi$ (in radian) for an 
incident electric vector
 $\Evi=(1,1)^T/\sqrt{2}$ and for the plate thickness $\ell_1$. The optical axis 
is parallel
 to the plane of interface.}
    \label{r}
\end{figure}

\begin{figure}[h]
\centering
\includegraphics{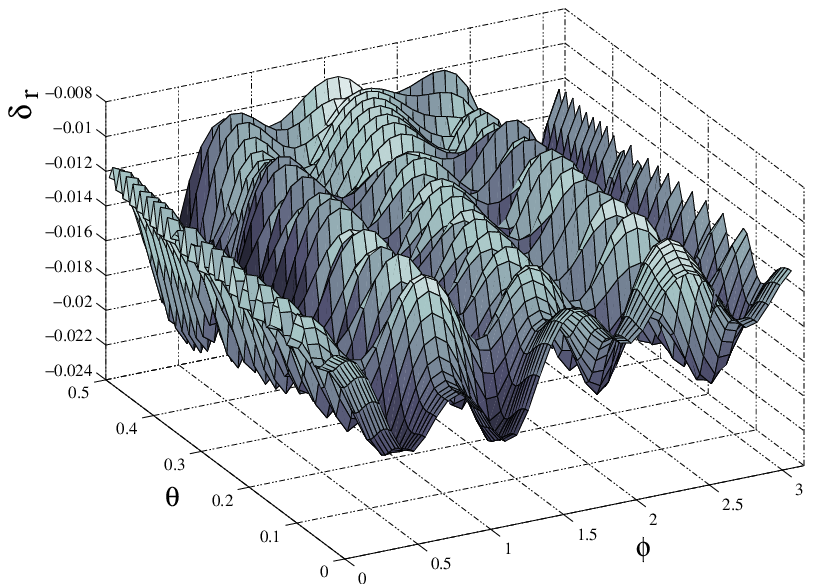}
\caption{$\delta r$ as a function of the angles $\theta$ and $\phi$ (in radian) 
for an incident electric vector
 $\Evi=(1,1)^T/\sqrt{2}$ and for the plate thickness $\ell_1$. The optical axis 
is parallel
 to the plane of interface.}
    \label{deltar}
\end{figure}

\begin{figure}[h]
\centering
\includegraphics{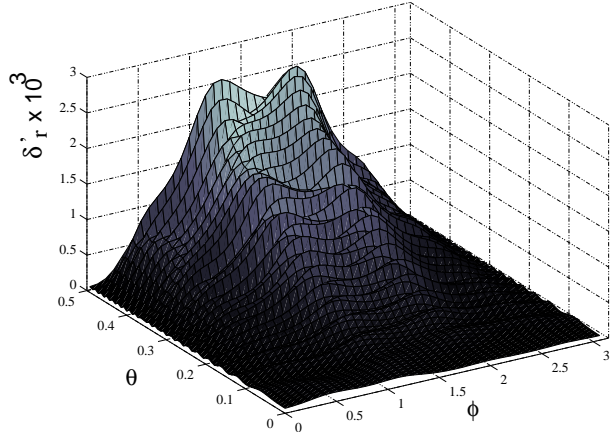}
\caption{$\delta' r$ as a function of the angles $\theta$ and $\phi$ (in radian) 
for an incident electric vector
 $\Evi=(1,1)^T/\sqrt{2}$ and for the plate thickness $\ell_1$. The optical axis 
is parallel
 to the plane of interface.}
    \label{deltarp}
\end{figure}

As mentioned in section \ref{oblique}, under normal incidence and for the 
 optical axis parallel to the plane of interface, the contribution of the 
optical
 activity to the platelet transmission matrix comes essentially from the 
coefficient
 $g_{11}$. The contribution of the coefficient $g_{33}$ appears under oblique 
incidence
 and can, {\it a priori}, be distinguished from the other one by considering 
various angles of
 incidence. We now consider the ratio $r=I_s/I_p$ where $I_s$ and $I_p$ are 
obtained by substituting
 $\shat$ and $\phat$ to $\xhat$ in Eq. (\ref{Ix-numeric}) respectively. To 
reproduce the 
 experimental results of Ref. \cite{chou}, we also fix $\psi=\pi/4$, {\it i.e.} 
$\Evi=(1,1)^T/\sqrt{2}$.
 The ratio $r$ is shown as a function of $\phi$ and $\theta$ for the plate 
thickness $\ell_1$
 in Fig. \ref{r}. To distinguish the contributions 
 from $g_{11}$ and $g_{33}$, we show in Figs. \ref{deltar} and \ref{deltarp}
 the relative differences $\delta r=(r-r_0)/r$ and $\delta' r=(r-r_1)/r$
 where $r_0$ and $r_1$ are obtained by setting $g_{11}=g_{33}=0$ and $g_{11}=0$ in the expression of $r$ respectively. From these figures, one sees that, as reported in  Ref. \cite{chou}, the contribution of the optical activity reach the percent level for certain values of $\phi$.


\clearpage

 In addition, we observe here that 
the contribution of 
  $g_{33}$ is negligible under normal incidence but reaches the few $10^{-3}$ 
level
  for reasonable values of the angle of incidence. The shape of the contribution 
of $g_{11}$ and $g_{33}$
 being different, it seems possible to determine both coefficients by measuring 
$r$ as a function of $\phi$
 for various angles of incidence. 

\begin{figure}[h]
\centering
\includegraphics{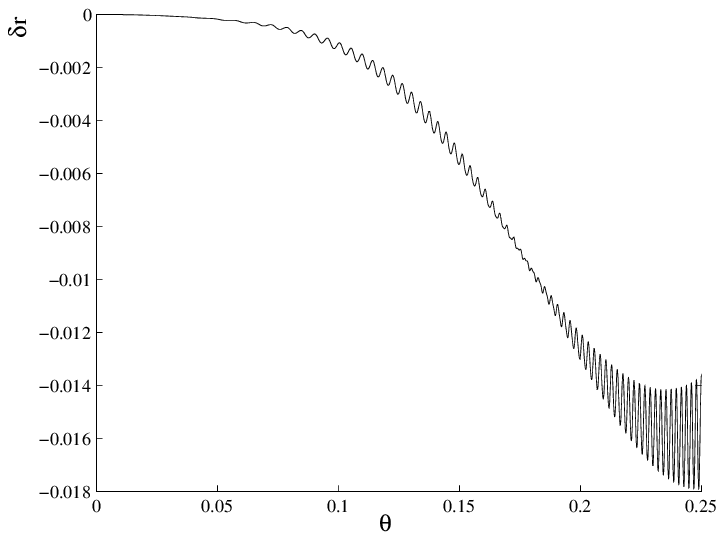}
\caption{$\delta r$ as a function of the angle $\theta$ (in radian) for an 
incident electric vector
 $\Evi=(1,1)^T/\sqrt{2}$ and for the plate thickness $\ell_5$. The optical axis 
is perpendicular
 to the plane of interface.}
    \label{deltar2}
\end{figure}

\begin{figure}[h]
\centering
\includegraphics{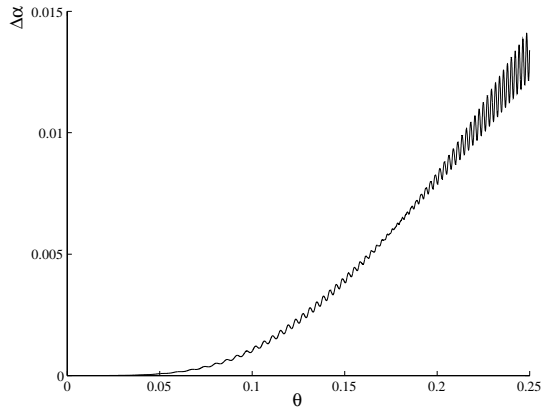}
\caption{$\Delta\alpha$, in radians, as a function of the angle $\theta$ (in 
radian)for an incident electric vector $\Evi=(1,1)^T/\sqrt{2}$ and for the plate thickness $\ell_5$. The optical axis 
is perpendicular to the plane of interface.}
    \label{phase}
\end{figure}

We finally consider the case of an optical axis perpendicular to the plane of 
interface. 
 We choose $\ell_5=1000\mu$m for the plate thickness
and, as previously, $\Evi=(1,1)^T/\sqrt{2}$ for the incident electric vector. In 
addition to
 $\delta r$, we introduce the phase angle $\alpha$ defined by the
 argument of the complex number $(\Evo\cdot\shat)/(\Evo\cdot\phat)$ and the 
 difference $\Delta\alpha=\alpha-\alpha_0$ where $\alpha_0$ is obtained 
 by setting
 $g_{11}=0$ in the expression of $\Evo$. Figs. \ref{deltar2} and \ref{phase} 
show
 $\delta r$ and $\Delta\alpha$ as a function of the angle of incidence $\theta$.

\clearpage
\noindent


\noindent
As in the case
 of an optical axis parallel to the plane of interface, the measurements of $r$ 
and $\Delta\alpha$
 as function of the angle of incidence lead to the determination of both 
$g_{11}$ and $g_{33}$. The 
 sensitivity to these parameters is even higher when the optical axis is 
perpendicular to the 
 plane of interface.

\section{Conclusion}

The transmission matrix for a uniaxial and chiral platelet has been calculated
 for an optical axis parallel and perpendicular to the plane of interface. The
 optical activity has been taken into account by considering the electric-
quadrupole and
 magnetic-dipole responses of the crystal medium.
 Simple and usable expressions have been derived for normal incidence and for 
the optical
 axis parallel and perpendicular to the plane of interface. The case of an 
oblique incidence 
 has been treated partly analytically and partly numerically.

 As an application, numerical studies of the 
 transmission of quartz platelets of various thicknesses have been provided.
 Under normal incidence, it was shown that the contribution of the
 optical activity can reach the percent level for quarter wave plates, 
independently of the 
 plate's order. Such a high contribution was found to be due
 to the transmission and reflection interface matrices. The case of half wave 
plates was also 
 investigated and, unlike quarter wave plates, the main contribution of the 
optical activity 
 came from the phase shift induced by the optical path inside the crystal. The 
variation of the 
 intensity after a quartz platelet was studied as a function of the angle of 
incidence and of the 
 orientation of the optical axis. A sensitivity to both gyration coefficients 
$g_{11}$ and $g_{33}$
 was observed under oblique incidence. This suggests that varying the angle of 
incidence one could
 determine experimentally these two 
 coefficients with a unique crystal cut. With respect to the results of Ref. 
\cite{bretagne},
  oblique incidence also offers the possibility to determine very accurately the 
platelet thickness and
 thereby the gyration coefficients.

\section*{Acknowledgement}

I would like to thank H. Feumi-Jentou and P.H. Burgat-Charvillon for their 
support.
 I would also like to thank F. Marechal for careful reading.


\end{document}